\documentclass[pra,twocolumn,showpacs,amsmath,amssymb,superscriptaddress]{revtex4}
\usepackage{graphicx}% Include figure files
\usepackage{color}
\usepackage{ulem}

\begin{document}

\title{Atypical BCS-BEC crossover induced by quantum-size effects}

\author{A. A. Shanenko}
\affiliation{Departement Fysica, Universiteit Antwerpen,
Groenenborgerlaan 171, B-2020 Antwerpen, Belgium}
\author{M. D. Croitoru}
\affiliation{Condensed Matter Theory Group, CPMOH, University of
Bordeaux I, France}
\author{A. V. Vagov}
\affiliation{University of Bayreuth, Institute of Theoretical
Physics III, D-95440 Bayreuth, Germany}
\author{V. M. Axt}
\affiliation{University of Bayreuth, Institute of Theoretical
Physics III, D-95440 Bayreuth, Germany}
\author{A. Perali}
\affiliation{School of Pharmacy, Physics Unit, University of Camerino,
I-62032-Camerino, Italy}
\author{F. M. Peeters}
\affiliation{Departement Fysica, Universiteit Antwerpen,
Groenenborgerlaan 171, B-2020 Antwerpen, Belgium}

\date{\today}
\begin{abstract}
Quantum-size oscillations of the basic physical characteristics of a confined fermionic condensate are a well-known phenomenon. Its conventional understanding is based on the single-particle physics, whereby the oscillations follow variations in the single-particle density of states driven by the size quantization. Here we present a study of a cigar-shaped ultracold superfluid Fermi gas, which demonstrates an important many-body aspect of the quantum-size coherent effects, overlooked previously. The many-body physics is revealed here in the atypical crossover from the Bardeen-Cooper-Schrieffer (BCS) superfluid to the Bose-Einstein condensate (BEC) induced by the size quantization of the particle motion. The single-particle energy spectrum for the transverse dimensions is tightly bound, whereas for the longitudinal direction it resembles a quasi-free dispersion. This results in the formation of a series of single-particle subbands (shells) so that the aggregate fermionic condensate becomes a coherent mixture of subband condensates. Each time when the lower edge of a subband crosses the chemical potential, the BCS-BEC crossover is approached in this subband, and the aggregate condensate contains both BCS and BEC-like components.
\end{abstract}


\maketitle

\section{Introduction}
\label{sec:intro}

A crossover from the weakly-interacting BCS superfluid of Cooper pairs to a BEC of tightly-bound molecule-like pairs~\cite{eagles,leggett} is one of the most important phenomena in the physics of fermionic condensates~(see also the review~\cite{bloch}). Although the BCS-BEC crossover was originally discussed in the context of semiconductor materials in the presence of superconducting correlations~\cite{eagles}~(for recent activity on this subject, see \cite{littlewood,neilson}), it was first demonstrated in experiments with ultracold superfluid Fermi gases~\cite{bloch}. In the standard scenario the BCS-BEC crossover for ultracold fermions is achieved via a Feshbach resonance in the particle scattering. Here we report a different mechanism to achieve the BCS-BEC crossover, inducing it by {\it quantum-size} (QS) effects in a cigar-shaped superfluid Fermi gas.

The QS effects in superfluid or superconducting systems have their origin in the geometric quantization of the single-particle motion and reveal themselves in the oscillations of the basic condensate properties. Such oscillations have been theoretically investigated in many different systems e.g., ultrathin films~\cite{blatt}; quantum striped superconductors and superconducting heterostructures at the atomic limit~\cite{perali,innocenti}; superconducting metallic nanowires~\cite{shan}; and a pancake-shaped superfluid Fermi gas~\cite{torma}. Recently, atomically uniform ${\rm Pb}$ nanofilms were fabricated, which resulted in the first experimental observation of the QS oscillations in the critical temperature~\cite{guo}. An interesting experimental study of $^6{\rm Li}$ Fermi gas in a pancake-shaped trap was also recently reported~\cite{dyke} where the effects of the transverse quantization on the aspect ratio of the atomic cloud were demonstrated. This opens new prospects for the study of the QS oscillations of the properties of fermionic condensates in ultracold Fermi gases with tunable confinement parameters. Conventional understanding of the QS oscillations follows from the fact that the single-particle energy spectrum for the quantum-confined dimensions is tightly bound, whereas along the other dimensions it resembles a quasi-free dispersion. This results in the formation of a series of single-particle subbands. The lower edge (bottom) of such a subband coincides with the corresponding discrete single-particle level associated with the confined dimensions. When the energy spacing between subbands is systematically decreased, e.g., by decreasing the relevant trapping frequency for superfluid fermionic atoms or the thickness of an atomically-uniform metallic nanofilm, the subband lower edges sequentially cross the chemical potential $\mu$. Each time this happens, the single-particle density of states (DOS) at $\mu$ increases, leading to a higher critical temperature, larger excitation gap, etc., which is referred to as the shape or QS resonance~\cite{blatt,perali,shan}.

In the present work, based on a study of a cigar-shaped ultracold superfluid Fermi gas, we demonstrate that the QS coherent effects cannot be fully understood in terms of the single-particle physics. The total or aggregate fermionic condensate in the system of interest is a coherent mixture of the subband components (condensates), and {\it each component} undergoes a BCS-BEC crossover when the lower edge of the corresponding single-particle subband crosses $\mu$. As a result of such an {\it atypical} BCS-BEC crossover, the total condensate is a coherent mixture of both the BCS and BEC-like components, which is most pronounced at the shape resonances.

This paper is organized as follows. In Sec.~\ref{sec:form} we outline the relevant formalism for calculating the ``wave function" of a condensed fermionic pair in the cigar-shaped superfluid Fermi gas. In Sec.~\ref{sec:cross} we analyze numerical results for the fermionic-pair ``wave function" with the focus on the atypical BCS-BEC crossover induced by the QS effects as dependent on the perpendicular (transverse) trapping frequency. In Sec.~\ref{sec:disc} we discuss technical details of the approximations used in our approach and show that the results here reported are not sensitive to these approximations. Our conclusions are given in Sec.~\ref{sec:conc}.

\section{Formalism}
\label{sec:form}

Our analysis is based on a numerical solution of the Bogoliubov-de Gennes (BdG) equations for $^6{\rm Li}$ atoms trapped by a harmonic axially symmetric potential $U({\bf r}) = M(\omega_{\perp}^2 \rho^2+\omega_{\parallel}^2 \,z^2)/2$, with ${\bf r}=\{\rho,\phi,z\}$ cylindrical coordinates and $\omega_{\parallel}\ll \omega_{\perp}$. Calculations are performed for zero temperature as the BdG equations are appropriate to describe the BCS-BEC crossover in spatially nonuniform fermionic systems at nearly zero temperatures. In particular, it has been shown in \cite{pieri} that the BdG equations reproduce the Gross-Pitaevskii equation for the condensate wave function on the BEC side of the crossover.

To clearly display the underlying physics, we employ the Anderson semi-analytical approximation~\cite{and} according to which the spatial dependence of the particle-like $u_{\nu}({\bf r})$ and hole-like $v_{\nu}({\bf r})$ wave functions appearing in the BdG equations is chosen to be proportional to the corresponding single-particle wave function $\varphi_{\nu}({\bf r})$~(see the discussion in Sec.~\ref{sec:disc}), i.e.,
\begin{equation}
u_{\nu}({\bf r})={\cal U}_{\nu}\varphi_{\nu}({\bf r}),\quad v_{\nu}
({\bf r})= {\cal V}_{\nu}\varphi_{\nu}({\bf r}),
\label{ander}
\end{equation}
where $\nu =\{j,n,m\}$ are the three quantum numbers associated with the longitudinal motion along the $z$ axis, the radial motion and the angular momentum, respectively. $\varphi_{\nu}({\bf r})=\vartheta_{nm}(\rho,\phi)\chi_j(z)$ is the product of the eigenfunctions of the 2D and 1D harmonic oscillators. Inserting Eq.~(\ref{ander}) into the BdG equations, one gets a system of two linear equations for the coefficients ${\cal U}_{\nu}$ and ${\cal V}_{\nu}$~(chosen real). A nontrivial solution exists when the corresponding determinant is equal to zero, which gives the quasiparticle energy $E_{\nu} =\sqrt{\lambda^2_{\nu} + \Delta_{\nu}^2}$, where $\lambda_{\nu}=\hbar\omega_{\perp}(1+2n+|m|) + \hbar\omega_{\parallel} (j+1/2) -\mu$ is the single-particle energy measured from the chemical potential, and $\Delta_{\nu}$ is the corresponding pairing energy. Then, together with the normalization condition ${\cal U}^2_{\nu}+{\cal V}^2_{\nu}=1$, the BdG equations yield ${\cal U}^2_{\nu} = \big(1 +\lambda_{\nu} /E_{\nu}\big)/2,\;{\cal V}^2_{\nu}=\big(1- \lambda_{\nu}/E_{\nu}\big)/2$. These expressions for ${\cal U}_{\nu}$ and ${\cal V}_{\nu}$ make it possible to find the BCS-like self-consistency equation given by (at $T=0$)
\begin{equation}
\Delta_{{\nu}}=\frac{1}{2}\sum\limits_{{\nu}^\prime} V_{{\nu}{\nu}^\prime}\,\Delta_{{\nu}^\prime}\,
\Big(\frac{1}{E_{{\nu}^\prime}} - \frac{1}{\lambda_{{\nu}^\prime}}\Big),
\label{3D}
\end{equation}
with the interaction matrix $V_{{\nu}{\nu}^\prime} = g\int {\rm d}^3r |\varphi_{\nu}({\bf r})|^2|\varphi_{{\nu}^\prime}({\bf r})|^2$ and $g$ the coupling constant. The second term in brackets of Eq.~(\ref{3D}) eliminates the ultraviolet divergence: this is a convenient simplification that, to first approximation, models the rigorous regularization for a spatially nonuniform system reported in \cite{bruun} (see the discussion in Sec.~\ref{sec:disc}). We avoid the ultra-confinement regime where the effective dimensionality of the system reduces and the particle scattering becomes different from the 3D case (see, e.g., \cite{bloch}). Here we take $\mu \gtrsim 2\hbar\omega_{\perp}$ and, in addition, the absolute value of the $s$-wave scattering length $a\,(a < 0)$ is chosen smaller than $l_{\parallel}, l_{\perp}$, where $l_{\alpha} = \sqrt{\hbar/(M \omega_{\alpha})}$. With this choice the interatomic collisions can be regarded as a 3D process (see the discussion in Sec.~\ref{sec:disc}) for which we use the standard expression of the pseudopotential theory $g=4\pi\hbar^2|a|/M$, with $M$ the atomic mass.

Equation~(\ref{3D}) can be viewed as a system consisting of multiple condensates with the pairing gaps $\Delta_{\nu}$ coupled through the interaction matrix $V_{\nu \nu^\prime}$. Furthermore, for the cigar-shaped trap the interlevel energy spacing corresponding to the quantization in the $z$ direction is sufficiently small so that the single-particle spectrum can be viewed as a sequence of the subbands $(n,m)$. Indeed, differences between $\Delta_\nu$'s within the same subband are almost insignificant (and disappear in the limit $l_\parallel \rightarrow \infty$). It is therefore useful to distinguish separate subband contributions to the system characteristics, i.e., to treat the system as a coherent mixture of multiple subband-dependent pair condensates.

By solving Eq.~(\ref{3D}) we obtain the set of $\Delta_\nu$. To probe the spatial pairing correlations (the main point of our study), the anomalous correlation function $\Psi({\bf r},{\bf r}^\prime)= \langle \hat{\psi}_{\uparrow}({\bf r})\hat{\psi}_{\downarrow} ({\bf r}^\prime) \rangle$ need to be calculated. Following the original works of Gor'kov\cite{gor} and Bogoliubov~\cite{bog}, it can be viewed as the wave function of a condensed fermionic pair. Using our subband-based classification, we can represent $\Psi({\bf r},{\bf r}^\prime)$~(using the Bogoliubov canonical transformation) as a sum over the relevant subbands
\begin{equation}
\Psi({\bf r},{\bf r}^\prime)=\sum\limits_{nm}\; \Psi_{nm} ({\bf r},{\bf r}^\prime),\label{Psi}
\end{equation}
where $\Psi_{nm}({\bf r},{\bf r}')=\vartheta_{nm}(\rho,\phi) \vartheta^*_{nm}(\rho',\phi')\;\psi_{nm}(z,z')$ and at zero temperature
\begin{align}
\psi_{nm}(z,z')=&\frac{1}{2}\sum\limits_j\,\chi_j(z)\;\chi^*_j(z')\,
\Delta_{nmj}\notag \\
&\times\Big(\frac{1}{E_{nmj}}-\frac{1}{\lambda_{nmj}}\Big).
\label{psiz}
\end{align}
Due to pairing, $\Psi({\bf r},{\bf r}^\prime)$ is localized as a function of the longitudinal relative coordinate $z-z^{\prime}$ (in the $x,y$ plane it is confined by the trapping potential) and the characteristic localization length, i.e., the longitudinal fermionic-pair size, is calculated as
\begin{equation}
\xi_0 = \Big({\cal N}^{-1}\int\!\!{\rm d}^3r {\rm d}^3r^\prime |\Psi({\bf r},{\bf r}^\prime)|^2(z-z')^2\Big)^{1/2},
\label{xi0}
\end{equation}
where ${\cal N} = \int\!{\rm d}^3r{\rm d}^3r^\prime|\Psi({\bf r},{\bf r}^\prime)|^2$ is the normalization factor. Similarly, one can define a subband-dependent fermionic-pair size $\xi^{(nm)}_0$ given by Eq.~(\ref{xi0}), with $\Psi({\bf r},{\bf r}^\prime)$ and ${\cal N}$ replaced by $\Psi_{nm}({\bf r},{\bf r}^\prime)$ and the corresponding normalization factor ${\cal N}_{nm}$. We note that while other definitions of the condensed-pair size are possible, the resulting expressions differ only by a constant factor~\cite{duk,ket}.

\section{Atypical BCS-BEC crossover}
\label{sec:cross}

As an illustration we consider a mixture of $^6{\rm Li}$ fermionic atoms with two interacting spin states, $|F,m_F\rangle = |1/2,1/2\rangle$ and $|1/2,-1/2\rangle$. The corresponding scattering length $a$ can be significantly modified through a broad Feshbach resonance with experimentally reported values on the BCS side from $-250\,{\rm nm}$ to $-100\,{\rm nm}$~\cite{sol}. For our numerical calculations we take $a=-140\,{\rm nm}$ and $-180\,{\rm nm}$. We choose $\omega_{||}/(2\pi)= 240\,{\rm Hz}$, $\mu = 100 \hbar\omega_{\parallel}$, and $\omega_\perp$ is assumed to be variable leading to tunable QS effects. (We found it convenient to present our results versus the ratio $s=\mu/\hbar\omega_{\perp}$). For the above choice of the parameters the particle density in the center of the trap is found to be about $10^{12}$-$10^{13}\,{\rm cm}^{-3}$, with $k_F|a|= 0.7$-$0.9$. This is consistent with most experiments where the particle density is reported in the range $10^{12}$-$10^{15}\,{\rm cm}^{-3}$ and $k_F|a|\sim 1$~\cite{bloch}. The characteristics of our trapping potential are chosen similar to those realized in recent experiments with quasi-1D Fermi gases. In particular, one can compare $\omega^{\rm exp}_{\parallel}/(2\pi) = 200\,{\rm Hz}$ and the ratio $\omega^{\rm exp}_{\parallel}/\omega^{\rm exp}_{\perp}=0.001$ reported in \cite{hulet} with our values $240\,{\rm Hz}$ and $0.02$-$0.04$. Note that $\omega_{\parallel} /\omega_{\perp}$ is larger in our calculations because, as already mentioned above, we avoid the ultra-confinement regime for the transverse motion of atoms
that is realized in \cite{hulet} and where only one transverse level (i.e., one subband) is occupied.

%
%%%%%%%%%%%%%%%%%%%%%%%%%%%%%%%%%%%%%%%%%%%%%%%%%%%%%%%%%%%%%%%%%%%%%%%%%%%%%%%%%%%%
\begin{figure}
\includegraphics[width=1.0\linewidth]{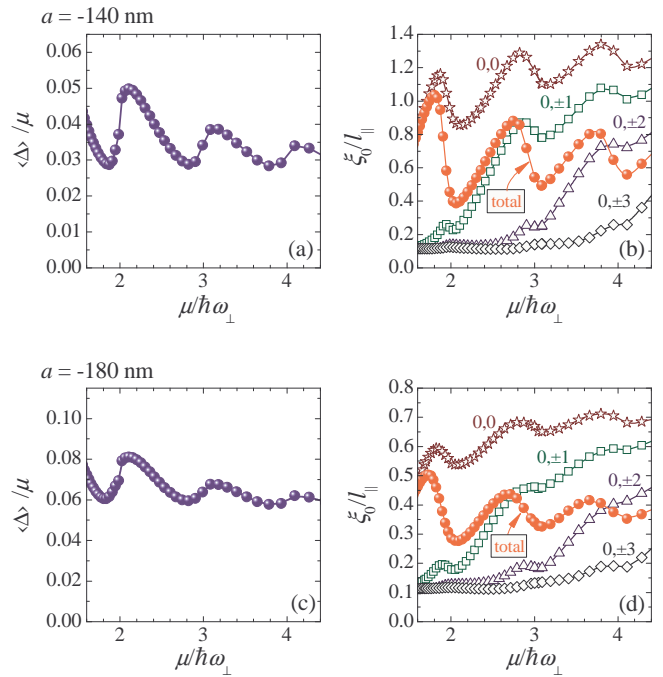}
\caption{(Color online) Panels (a) and (c) demonstrate the QS oscillations of the averaged pairing gap $\langle \Delta \rangle$ for scattering lengths $a=-140\,{\rm nm}$ and $-180\,{\rm nm}$, respectively. Panels (b) and (d) show the corresponding changes in the longitudinal fermionic-pair size $\xi_0$~[see Eq.~(\ref{xi0})] and the subband fermionic-pair size $\xi^{(nm)}_0$ for subbands $(n,m)=(0,0),\,(0,\pm1)$ and $(0,\pm2)$.} \label{fig1}
\end{figure}
%%%%%%%%%%%%%%%%%%%%%%%%%%%%%%%%%%%%%%%%%%%%%%%%%%%%%%%%%%%%%%%%%%%%%%%%%%%%%%%%%%%
%

\subsection{Oscillations of longitudinal fermionic-pair size}
\label{sec:qs}

To reveal the many-body aspect of the QS effects, we first need to consider how $\Delta_{\nu}$'s change with $s=\mu/\hbar\omega_{\perp}$. Figures~\ref{fig1}(a) and (c) show $\langle\Delta\rangle$ i.e., the pairing gap averaged over the Fermi-surface, as a function of $s$ calculated for $a=-140\,{\rm nm}$ and $-180 \,{\rm nm}$, respectively. As can be seen, $\langle\Delta\rangle$ increases in the vicinity of integer values of $s$. This condition for the developing of a shape resonance~\cite{blatt} is satisfied when the bottom of a subband, referred to as the resonant subband, approaches the chemical potential and the DOS increases. As a result, $\langle\Delta\rangle$ exhibits QS oscillations with changing $\omega_{\perp}$ similar to those reported for a pancake-shaped superfluid Fermi gas~\cite{torma}.

Now, based on the data for $\langle\Delta\rangle$, we demonstrate that the conventional single-particle picture of the QS oscillations fails to explain the corresponding changes in the two-particle characteristics. This is illustrated by the results for the longitudinal fermionic-pair size $\xi_0$ shown in Figs.~\ref{fig1}(b) and (d). Similar to the averaged pairing gap, $\xi_0$ exhibits remarkable QS oscillations. However, using the standard BCS estimate $\xi_0 \propto \hbar v_F/\langle\Delta\rangle$, with $v_F$ the Fermi velocity at the center of the trap, and taking into account the variations of $\langle \Delta \rangle$ in Fig.~\ref{fig1}, we obtain for $\xi_0$ a decrease by a factor of $1.6$ when $s$ increases from $1.8$ to $2.1$ at $a=-140\,{\rm nm}$. Note that the corresponding change in $v_F$ is negligible, see the data for single-particle density discussed in Sec.~\ref{sec:agg}. This is a considerable underestimation of the numerical results given by Fig.~\ref{fig1}(b), where $\xi_0|_{s=1.8}/\xi_0|_{s=2.1}\approx 3$. A similar discrepancy is found for $a=-180\,{\rm nm}$.

A detailed analysis shows that this discrepancy is related to a significant redistribution of the fermionic condensate over the available subbands. This is seen from Fig.~\ref{fig1}(b), where $\xi_0$ is compared with $\xi_0^{(nm)}$ for subbands $(n,m)=(0,0),\,(0,\pm1),\,(0,\pm2)$. When $s \le 1.9$, the main contribution to the pair condensate comes from subband $(n,m)=(0,0)$, i.e., $\Psi \approx \Psi_{0,0}$, and we obtain $\xi_0 \approx \xi_0^{(0,0)}$. As the system goes through the resonance that develops at $s=2$, $\xi_0$ drops and approaches $\xi_0^{(0,\pm 1)}$. In this case, two resonant subbands $(n,m)=(0,\pm1)$ make the largest contribution to the total condensate i.e., about $70\%$ at $s=2.1$. At the next resonance, $s = 3$, $\xi_0$ decreases again and approaches $\xi_0^{(0,\pm 2)}$, which points to the enhancement of the contribution $\Psi_{0,\pm 2}$. At larger $s$ the effect is weakened because the total number of contributing subbands increases while the relative contribution of a particular resonant subband diminishes. Results for $a=-180\,{\rm nm}$ in Fig.~\ref{fig1}(d) exhibit a similar behavior.

%
%%%%%%%%%%%%%%%%%%%%%%%%%%%%%%%%%%%%%%%%%%%%%%%%%%%%%%%%%%%%%%%%%%%%%%%%%%%%%%%%%%%
\begin{figure}
\includegraphics[width=1.0\linewidth]{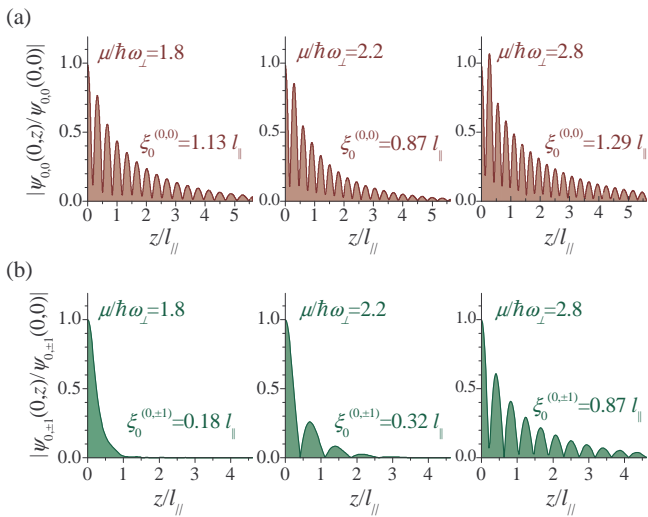}
\caption{(Color online) Spatial profile of the modulus of $\psi_{nm}(0,z)$ for subbands $(0,0)$~(a) and $(0,\pm1)$~(b). The data for $\mu/\hbar\omega_{\parallel} = 1.8,\,2.2$ and $2.8$ are given in each panel and calculated at zero temperature.}
\label{fig2}
\end{figure}
%%%%%%%%%%%%%%%%%%%%%%%%%%%%%%%%%%%%%%%%%%%%%%%%%%%%%%%%%%%%%%%%%%%%%%%%%%%%%%%%%%%%
%

\subsection{Squeezing of fermionic-pair "wave function"}
\label{sec:squeez}

The arguments given in Sec.~\ref{sec:qs} demonstrate that the redistribution of
the fermionic condensate over the available subbands has a significant effect on $\xi_0$. Another contributing factor is the large variation in the subband pair size $\xi_0^{(n,m)}$ when the bottom of the corresponding subband crosses $\mu$, as seen in Figs.~\ref{fig1}(b) and (d). A further insight is obtained by considering how  $\Psi_{nm}({\bf r},{\bf r}^{\prime})$ decays with increasing $z-z^{\prime}$. This decay of the subband pair wave-function $\Psi_{nm}({\bf r},{\bf r}^{\prime})$ is controlled by its longitudinal component $\psi_{nm} (z,z^{\prime})$ defined by Eq.~(\ref{psiz}). Figure~\ref{fig2} shows $|\psi_{nm}(0,z)|$ as a function of $z$ for subbands $(n,m)=(0,0)$~[panel (a)] and $(0,\pm1)$~[panel (b)] at $s=1.8,\,2.2$ and $2.8$~(for $a=-140\,{\rm nm}$). As seen, $\psi_{0,0}(0,z)$ is a slowly decaying oscillatory function for all values of $s$, which is typical for loosely-bound Cooper pairs in a bulk superconductor. Contrary to this, $\psi_{0,\pm1}(0,z)$ exhibits a crossover from the strongly localized (at $s=1.8$) to the BCS weakly localized regime (at $s=2.8$) when the system passes through the resonance associated with $s=2$. At $s=1.8$ the lower edge of subbands $(n,m)=(0,\pm1)$ is situated slightly above $\mu$, whereas at $s=2.2$ and $s=2.8$ it is slightly below and far below $\mu$, respectively. The edge of subband $(0,0)$ is far below $\mu$ for all given values of $s$. Notice that at $s = 1.8$ subband $(0,0)$ makes a contribution of about $95\%$ to the total condensate while subbands $(0,\pm1)$ yield only $5\%$. At $s=2.2$ the contribution of subband $(0,0)$ decreases down to $30\%$, as opposed to the contribution of subbands $(0,\pm1)$ that increases up to about $70\%$.

The results of Fig.~\ref{fig2} are understood as follows. When the lower edge of a single-particle subband is far below $\mu$, the ratio of the pair-interaction energy~\cite{note} to the longitudinal kinetic energy in this subband is small, as expected in conventional weak-coupling BCS theory. However, when the subband edge approaches $\mu$, the longitudinal kinetic energy is reduced, and the ratio strongly increases, which means that the pair-interaction prevails over the longitudinal motion. This forces the fermionic pairs in this subband to squeeze in the longitudinal direction [see Fig.~\ref{fig2}(b)]. Thus, the effect follows from a redistribution of the kinetic energy between the longitudinal and transverse degrees of freedom in the subband whose lower edge crosses $\mu$. This drop in the fermionic-pair size is in fact an atypical example of the BCS-BEC crossover that takes place in a single subband. However, unlike previously discussed systems, here the crossover is driven by the QS effects, which prompted the term {\it atypical}. Qualitative difference between the partial condensates associated with different subbands explains the failure of the BCS estimate $\xi_0 \propto \hbar v_F/\langle \Delta \rangle$ in the analysis of the results in Figs.~\ref{fig1}(b) and (d).

Since the classical paper by Cooper\cite{cooper} it is well-known that the configuration of the phase space available for the scattering of time-reversed fermions plays a crucial role for the formation of condensed fermionic pairs. Indeed, only a strong enough attractive interaction between fermions with opposite spin in 3D is able to produce a two-body bound state in the vacuum. However, when
the scattering of fermions is influenced by the presence of a filled Fermi sea, i.e., the available phase space is restricted by exclusion of the single-particle states inside the Fermi sea, we arrive at the Cooper instability resulting in the formation of weakly bound in-medium pairs of fermions for arbitrary strength of
the attractive interaction. Restricting the phase space by removing the filled Fermi sea, one actually removes long range contributions in the Cooper-pair wave function, which, say, ``encourage" fermions to form in-medium bound states. Our results show that the additional reconfiguration of the phase space, such that the band of single-particle states splits up into a series of lower-dimensional subbands, can further modify the scenario of pairing through the atypical BCS-BEC crossover.

\subsection{Coherent mixture of BCS and BEC-like condensates}
\label{sec:agg}

%
%%%%%%%%%%%%%%%%%%%%%%%%%%%%%%%%%%%%%%%%%%%%%%%%%%%%%%%%%%%%%%%%%%%%%%%%%%%%%
\begin{figure}
\includegraphics[width=0.85\linewidth]{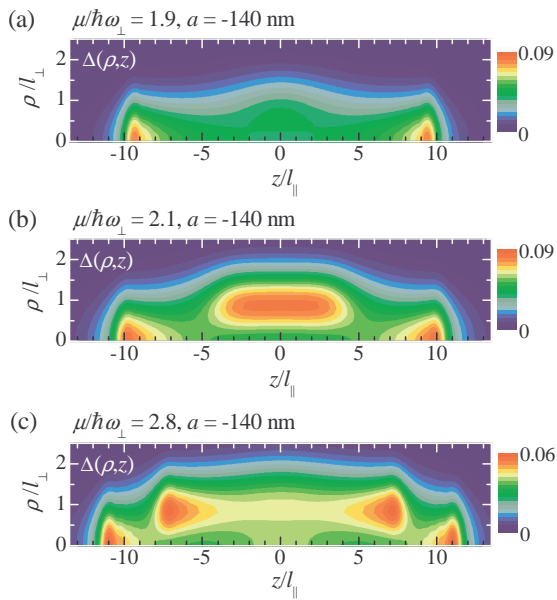}
\caption{(Color online) Contour plots of the order parameter $\Delta(\rho,z)$ obtained for $a=-140\,{\rm nm}$~(at $T=0$) below the resonance at $\mu/\hbar \omega_{\perp} =1.9$~(a), in the vicinity of the resonance  at $\mu/\hbar \omega_{\perp} =2.1$~(b) and far beyond it at $\mu/\hbar\omega_{\perp} =
2.8$~(c).}
\label{fig3}
\end{figure}
%%%%%%%%%%%%%%%%%%%%%%%%%%%%%%%%%%%%%%%%%%%%%%%%%%%%%%%%%%%%%%%%%%%%%%%%%%%%%
%

The BCS-BEC crossover in a resonant subband also reveals itself in the properties of the aggregate condensate of the system. However, the effect depends on the relative contribution of a resonant subband. It is maximal at the corresponding resonance and diminishes away from the resonance. This contribution similarly drops when the total number of relevant subbands increases, and thus, as for other quantities, the lowest resonances (i.e., for $s=2$ and $3$) are the most visible in $\xi_0$. Effects of the subband BCS-BEC crossover on the aggregate condensate can be estimated by calculating the quantity $\gamma=k_F \xi_0$~\cite{Pistolesi94}. For the BCS system $\gamma \gg 1$, i.e., the Cooper-pair size greatly exceeds the mean distances between particles, and this leads to a considerable overlap between the fermionic pairs. For $\gamma \ll 1$ such an overlap is absent, and the system becomes a BEC of tightly-bound point-like molecules. The intermediate crossover region is reached when $1/\pi\lesssim \gamma \lesssim 2\pi$~\cite{Pistolesi94}. For $s=1.9$ i.e., just before the resonance at $s=2$, we obtain $\gamma \approx 10$ and so the system is in the BCS regime. At $s=2.1$ we find $\gamma \approx 4$, which corresponds to the intermediate regime of the BCS-BEC crossover. The size-dependent drops in $\xi_0$ become larger for smaller $|a|$~(on the BCS side of the Feshbach resonance), see Fig.~\ref{fig1}. The reason for this is twofold. First, at smaller $|a|$ the energy window for contributing subbands also decreases, which means that the relative contribution of resonant subbands is larger. Second, the difference between $\xi_0^{(nm)}$ in neighboring subbands increases, as seen in Figs.~\ref{fig1}(c) and (d), and this increases the magnitude of variations in $\xi_0$.

Variations in $\xi_0$ are accompanied by substantial changes in the spatial profile of the order-parameter $\Delta({\bf r})=-g\Psi({\bf r},{\bf r})$. Figure~\ref{fig3} shows the contour plots of $\Delta(\rho,z)$ for $a = -140\,{\rm nm}$ and $T=0$, calculated (a) slightly below the resonance, at $s=1.9$; (b) close to the resonance, at $s=2.1$, and (c) above the resonance, at $s=2.8$. In the first case the pair condensate is almost uniformly distributed over the trap with two peaks at the edges of the condensate cloud. These peaks are typical for a confined BCS condensate and can be explained by the presence of the turning points in the trajectories of particles with energy close to $\mu$. When the resonance develops [Fig.~\ref{fig3}(b)], the spatial distribution of the condensate acquires a pronounced bimodal character with an additional sizeable peak around $z=0$ due to the contribution of resonant subbands $(0,\pm1)$. The bimodal character clearly indicates that the system becomes a coherent mixture of two qualitatively different condensates: the first is associated with subband $(0,0)$ and has properties typical for the BCS system; the second is due to resonant subbands $(0,\pm1)$ and displays a typical BEC-like behavior (the formation of bosonic-like states distributed at the center of the trap).

%
%%%%%%%%%%%%%%%%%%%%%%%%%%%%%%%%%%%%%%%%%%%%%%%%%%%%%%%%%%%%%%%%%%%%%%%%%%%%%%%%%%%
\begin{figure}
\includegraphics[width=0.85\linewidth]{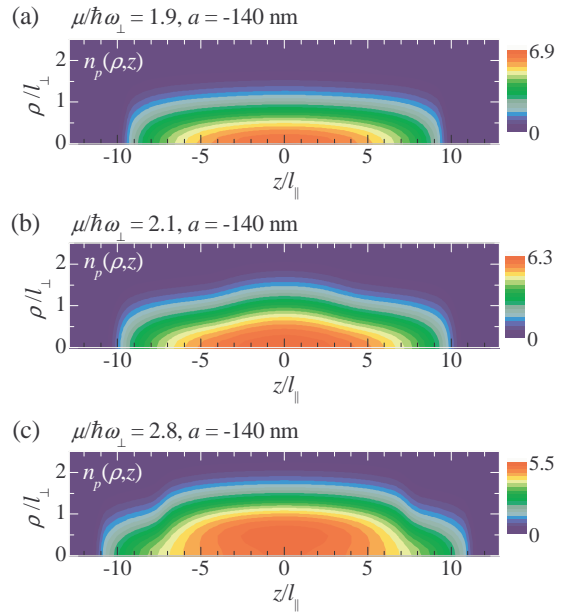}
\caption{(Color online) Contour plots of the single-particle density $n_p(\rho,z)$ (given in units of $10^{12}\,{\rm cm}^{-3}$) calculated for the same parameters as in Fig.~\ref{fig3}(a)-(c), respectively.}
\label{fig4}
\end{figure}
%%%%%%%%%%%%%%%%%%%%%%%%%%%%%%%%%%%%%%%%%%%%%%%%%%%%%%%%%%%%%%%%%%%%%%%%%%%%%%%%%%%
%

Note that contrary to the order parameter, the spatial distribution of atoms does not exhibit such noticeable changes at resonances, as seen from the comparison of Fig.~\ref{fig3} with Fig.~\ref{fig4}, where contour plots of the position dependent single-particle density $n_p({\bf r})$ are given. Due to the axial symmetry, we have $n_p({\bf r})=n_p(\rho,z)$, and at zero temperature
\begin{equation}
n_p(\rho,z)=\sum\limits_{nmj}\left(1-\frac{\lambda_{nmj}-\mu}{E_{nmj}}\right)
|\vartheta_{nm}(\rho,\varphi)|^2\,\chi^2_j(z),
\label{eq:dens}
\end{equation}
where the absolute value of the eigenfunction $\vartheta_{nm} (\rho,\varphi)$ does not depend on $\varphi$. The sum in Eq.~(\ref{eq:dens}) is convergent and the ultraviolet regularization is not required.

\section{Discussion}
\label{sec:disc}

Here we discuss details of approximations involved in our study, including the Anderson solution to the BdG equations, the ultraviolet regularization, and treating the interatomic collisions as a 3D process.

{\itshape The Anderson solution}. Our calculations are based on the BdG equations that can be, in principle, solved without additional approximations. However, for the sake of transparent physical interpretations, here we employ Anderson's recipe for an approximate solution to the BdG equations, see Eq.~(\ref{ander}). We recall that the Anderson approximation incorporates only the pairing of the time-reversed states~\cite{and}, and the power of this approximation is based on the fact that the interaction matrix elements involving the time-reversed states are most pronounced as compared to other pairing combinations. Corrections to the Anderson solution were found in the range of a few percent for quasi-1D superconducting condensate in metallic nanowires~\cite{shan}, which allows one to expect a similar accuracy of the Anderson approximation for a cigar-shaped ultracold superfluid Fermi gas.

Notice that the energy spacing between the single-particle levels is not directly related to the validity of the Anderson ansatz. However, it can be an additional factor improving accuracy of the ansatz. Indeed, when the energy spacing between single-particle levels exceeds the characteristic pairing energy, the pairing of non-time-reversed states is suppressed not only due to smaller coupling but, in addition, due to a large interlevel spacing between the non-time-reversed states. In this case the Anderson approximation becomes practically exact. However, when the interlevel energy spacing is small as compared to the pairing energy, this does not necessarily mean a breakdown of the Anderson approximation. The pairing of the non-time-reversed states will not be pronounced anyway, thanks to the classical argumentation by Anderson~\cite{and}. For instance, Anderson's recipe yields exact solution for bulk superconductors where the pairing energy is much larger than the interlevel energy spacing.

%
%%%%%%%%%%%%%%%%%%%%%%%%%%%%%%%%%%%%%%%%%%%%%%%%%%%%%%%%%%%%%%%%%%%%%%%%%%%%%%%%%%%
\begin{figure}[t]
\resizebox{0.57\columnwidth}{!}{\rotatebox{0}{
\includegraphics{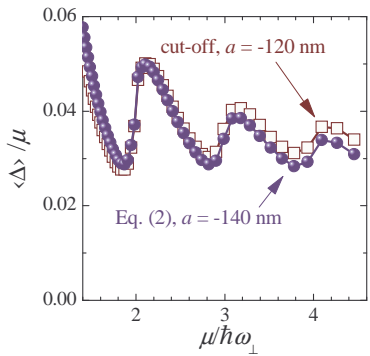}}}
\caption{The averaged energy gap $\langle\Delta\rangle$~(in units of $\mu$) versus $\mu/\hbar\omega_{\perp}$: squares are results for $a=-120\,{\rm nm}$ calculated with the Bruun-Heiselberg cut-off $|\lambda_{\nu}| < \mu$; circles are the data from Fig.~\ref{fig1}(a).} \label{fig5}
\end{figure}
%%%%%%%%%%%%%%%%%%%%%%%%%%%%%%%%%%%%%%%%%%%%%%%%%%%%%%%%%%%%%%%%%%%%%%%%%%%%%%%%%%%
%

{\itshape The ultraviolet regularization}. Notice that our simplified 3D-like regularization differs from the rigorous procedure reported by Bruun and coauthors \cite{bruun} for spatially nonuniform systems by the absence of an additional corrective term. However, following arguments by Bruun himself and Heiselberg in \cite{bruun1} [see the discussion just above Eq.~(15) in this reference], one can expect that the role of this corrective term is not significant. Moreover, the authors of Ref.~\onlinecite{bruun1} introduced a simple cut-off $|\lambda_{\nu}| < \mu$ and argued that such a cut-off is ``a first approximation to the more rigorous procedure of Bruun et al." for trapped Fermi gases. In Fig.~\ref{fig5} we show results for $\langle\Delta\rangle$ calculated with this cut-off versus our data from Fig.~\ref{fig1}(a). Though $\langle\Delta\rangle$ based on the cut-off procedure is slightly different as compared to the averaged gap calculated from Eq.~(\ref{3D}), this difference practically disappears with a small shift in the scattering length, i.e., $a \to a + 20\,{\rm nm}$, in the cut-off data. Thus, one can conclude that our simplified ultraviolet regularization, which becomes exact by increasing the number of subbands as the radial confinement is reduced, does not affect our conclusions.

{\itshape Interatomic collisions}. Several explanations about interatomic collisions in the cigar-shaped Fermi gas are also needed. For our trapping potential we have $\omega_{\parallel}/\omega_{\perp}\sim 0.01$, and looking at this aspect ratio, one might get the impression that the two-particle scattering has an effectively-1D character here. However, this is not correct. Though the character of the interatomic scattering is rather complex in quasi-1D systems, the choice of the 3D pseudopotential is well justified for our range of parameters.

Based on the results for the binary atomic collisions in the quasi-2D tightly confined system~\cite{petrov}, one can consider that the scattering amplitude in the quasi-1D system is a function of the two important parameters $|a|/l_{\perp}$ and $E_{\rm sc}/\hbar \omega_{\perp}$~(with $E_{\rm sc}$ the scattering energy). Pronounced deviations from the ordinary 3D scattering can be expected~\cite{petrov} when $|a|/l_{\perp}\gg 1$ and $E_{\rm sc}/\hbar \omega_{\perp} \ll 1$~(see also Refs.~\onlinecite{olsh} and \onlinecite{berg}).

In our calculations $l_{\perp}$ varies from $0.35{\rm \mu m}$ to $0.5{\rm \mu m}$ when $\mu/\hbar\omega_{\perp}$ increases from $2$ to $4$. So, one finds that $|a|/l_{\perp} \approx 0.3$-$0.4$ for $a=-140\,{\rm nm}$. The relevant scattering energies in our problem can be roughly estimated as twice the chemical potential measured from the lowest single-particle energy, i.e., $E_{\rm sc} \sim 2(\mu-\hbar\omega_{\perp})$. Using this estimate we find that for our parametric choice $E_{\rm sc}/\hbar\omega_{\perp} \gtrsim 2$~(as $\mu \gtrsim 2\hbar\omega_{\perp}$), which is directly related to the fact that our study is focused on the case of multiple contributing subbands. Thus, $E_{\rm sc}/\hbar\omega_{\perp}$ is too large and $|a|/l_{\perp}$ is too small to favor pronounced 1D-modifications to the interatomic collisions. We note that the choice of the 3D scattering length $a=-140\,{\rm nm}$ and $-180\,{\rm nm}$ is not crucial for our predictions of the atypical BCS-BEC crossover. We have performed additional calculations and found even more pronounced variations of the fermionic-pair size at $a=-100\,{\rm nm}$~(see the discussion in the first paragraph of Sec.~\ref{sec:agg}).

It is also important to note that there is one exception when the estimate of $E_{\rm sc}$ in the previous paragraph does not hold. In a resonant subband, whose bottom is located in the vicinity of the chemical potential, the longitudinal motion of atoms is depleted. As a result, the relevant energies of the interband scattering in such a subband can be smaller than $\hbar\omega_{\perp}$. In this case modifications to the pseudopotential could be pronounced, including the appearance of the confinement-induced Feshbach resonance~\cite{olsh,berg}, if the ratio $|a|/l_{\perp}$ were large, i.e.,  $|a|/l_{\perp}\gtrsim 1$. However, for $|a|/l_{\perp} \ll 1$~(this is our case) the 1D-modifications are reduced to almost insignificant renormalization of $a$, see Ref.~\onlinecite{olsh}.

Thus, we can conclude that the 1D-modifications to the 3D pseudopotential can be neglected for our choice of physical parameters.

\section{Conclusions}
\label{sec:conc}

In conclusion, we have demonstrated an atypical BCS-BEC crossover induced by the Quantum-Size effects for a $^6{\rm Li}$ superfluid gas in a cigar-shaped trap. For such a trap geometry the transverse quantization of the particle motion results in the formation of single-particle subbands so that the fermionic condensate becomes a coherent mixture of subband-dependent different condensates. Each time the lower edge of a subband crosses the chemical potential, the subband fermionic-pair size drops so that the fermionic pairing in this subband changes qualitatively, displaying the BCS-BEC crossover. As a result, the total fermionic condensate becomes a coherent mixture of BCS and BEC-like components, and the longitudinal pair size $\xi_0$ associated with the aggregate condensate decreases.
Radio-frequency spectroscopy can be used to detect the quantum-size driven squeezing of fermionic pairs~\cite{ket}. Note that a similar many-body physics driven by the Quantum-Size oscillations can be expected for a pancake-shaped superfluid Fermi gas with only a few available transverse levels (the experimental realization of such a system, with observation of many subbands due to quantum confinement, was recently reported in \cite{dyke}).

\begin{acknowledgments}

This work was supported by the Flemish Science Foundation (FWO-Vl). The authors thank C. Salomon and C. Vale for their valuable explications of the experimental situation and interest to our work. We are grateful to G. Strinati, D. Neilson, and P. Pieri for useful discussions. M.D.C. acknowledges support of the EU Marie Curie IEF Action (Grant Agreement No. PIEF-GA-2009-235486-ScQSR). A.P. gratefully acknowledges financial support of the European Science Foundation, POLATOM Research Networking Programme, Ref. n. 4844 for his visit to the University of Antwerp. A.A.S. acknowledges financial support of the European Science Foundation, POLATOM Research Networking Programme, Ref. n. 5200 for his visit to the University of Camerino.

\end{acknowledgments}


\begin{thebibliography}{99}
\bibitem{note} The averaged gap $\langle\Delta\rangle$ is a convenient measure of the pair-interaction energy.
\end{thebibliography}
\end{document}